\pdfoutput=1

\documentclass[11pt,a4paper]{article}
\usepackage{bm}
\usepackage{jheppub}
\usepackage{mathrsfs}

\let\de=\partial
\DeclareMathOperator{\tr}{tr}
\DeclareMathOperator{\sn}{sn}
\DeclareMathOperator{\dn}{dn}
\newcommand{\dd}{\text{d}}
\newcommand{\La}{\mathcal{L}}
\newcommand{\Ha}{\mathcal{H}}
\newcommand{\imag}{\text{i}}
\newcommand{\Ge}{\mathcal{G}}
\newcommand{\gr}[1]{\text{#1}}
\newcommand{\vek}[1]{\bm{#1}}
\allowdisplaybreaks

\title{Chiral soliton lattice in QCD-like theories}

\author{Tom\'a\v{s} Brauner,}
\author{Georgios Filios}
\author{and Helena Kole\v{s}ov\'{a}}
\affiliation{Department of Mathematics and Physics, University of Stavanger,\\
N-4036 Stavanger, Norway}
\emailAdd{tomas.brauner@uis.no}
\emailAdd{georgios.filios@uis.no}
\emailAdd{helena.kolesova@uis.no}

\abstract{Recently, it has been shown that the ground state of quantum chromodynamics (QCD) in sufficiently strong magnetic fields and at moderate baryon number chemical potential carries a crystalline condensate of neutral pions: the chiral soliton lattice (CSL)~\cite{Brauner:2016pko}. While the result was obtained in a model-independent manner using effective field theory techniques, its realization from first principles using lattice Monte Carlo simulation is hampered by the infamous sign problem. Here we show that CSL, or a similar inhomogeneous phase, also appears in the phase diagram of a class of vector-like gauge theories that do not suffer from the sign problem even in the presence of a baryon chemical potential and external magnetic field. We also show that the onset of nonuniform order manifests itself already in the adjacent homogeneous Bose-Einstein-condensation phase through a characteristic roton-like minimum in the dispersion relation of the lowest-lying quasiparticle mode. Last but not least, our work gives a class of explicit counterexamples to the long-standing conjecture that positivity of the determinant of the Dirac operator (that is, absence of the sign problem) in a vector-like gauge theory precludes spontaneous breaking of translational invariance, and thus implies the absence of inhomogeneous phases in the phase diagram of the theory.}


\begin{document}
 
\maketitle


\section{Introduction}
\label{sec:intro}

The concept of spontaneous symmetry breaking has been one of the most fruitful paradigms in modern quantum physics, underlying a broad range of fascinating phenomena from superfluidity to ferromagnetism and the existence of light pseudoscalar mesons in the hadron spectrum. Starting from the \emph{assumption} that the ground state of a given quantum system has lower symmetry than its equations of motion, one can take advantage of the powerful tools of effective field theory (EFT) to work out a model-independent description of the system's low-energy physics in terms of the ensuing Nambu-Goldstone (NG) bosons. The development of low-energy EFT for broken symmetries was pioneered half a century ago by Weinberg~\cite{Weinberg:1968de} and others~\cite{Coleman:1969sm,Callan:1969sn}, and its applications are still far from being exploited.

Much less is known about the conditions under which a continuous symmetry of a quantum system can, or \emph{cannot}, be spontaneously broken. With rare exceptions where absence of spontaneous breaking can be proven~\cite{Lieb:1961fr,Mermin:1966fe,Hohenberg:1967zz,Coleman:1973ci,Vafa:1983tf,Bruno:2013mva,Watanabe:2014hea}, one likewise simply assumes, based on experience or intuition, that the ground state of a certain system maintains the full symmetry of its dynamics. In this regard, an intriguing proposal was made in ref.~\cite{Splittorff:2000mm}, linking the absence of spontaneous breaking of spatial translations in vector-like gauge theories (hereafter simply QCD-like theories) to the positivity of the determinant of their Dirac operator. The latter condition is required in order for the standard importance sampling lattice Monte Carlo techniques to be applicable; its violation is usually referred to as the sign problem.\footnote{For the sake of brevity, we will use throughout the paper the term ``sign problem'' as a synonym for violation of the positivity condition on the determinant of the Dirac operator. The same term is sometimes used in a broader sense whereby the sign problem may or may not be present, or may be severe or mild, depending on the choice of variables in which the functional integral of the theory is numerically computed.} If true, the conjecture of ref.~\cite{Splittorff:2000mm} would thus make a remarkable addition to the list of existing no-go theorems for spontaneous symmetry breaking and at the same time, on the practical side, seriously hamper first-principle studies of inhomogeneous phases in dense quark matter.

The latter are not of mere academic interest. Namely, it is now widely believed that a substantial portion of the QCD phase diagram is occupied by inhomogeneous phases of various kinds, see ref.~\cite{Buballa:2014tba} for a review. Apart from its intrinsic interest, spontaneous formation of nonuniform order also has important consequences for the thermodynamics of dense nuclear/quark matter. However, a majority of the predictions of inhomogeneous phases in the QCD phase diagram is based on model calculations neglecting the effects of order parameter fluctuations, which may eventually render such phases unstable at any nonzero temperature~\cite{Baym:1982ca,Lee:2015bva,Hidaka:2015xza}. A rare exception of a phase whose prediction is model-independent and under theoretical control, is the chiral soliton lattice (CSL). This is a macroscopic manifestation of the chiral anomaly, whereby matter with nonzero baryon density can be realized without baryons, through a solitonic condensate of neutral pions. The CSL state requires for its existence either a sufficiently strong external magnetic field~\cite{Brauner:2016pko} or global rotation~\cite{Huang:2017pqe}.

The goal of this paper is to show that the CSL phase, or a similar inhomogeneous phase, also appears in the phase diagram of an infinite class of QCD-like theories free of the sign problem. This has two immediate consequences. First, we disprove the above-mentioned conjecture that absence of sign problem implies absence of inhomogeneous phases in the phase diagram~\cite{Splittorff:2000mm}. Second, our model-independent prediction of the CSL phase in theories amenable to lattice Monte Carlo simulations might provide a useful test bed for future numerical studies of dense quark matter. (To the best of our knowledge, the first attempt to study a crystalline phase of baryonic matter in lattice simulation was made in ref.~\cite{deForcrand:2006zz}, based on the $(1+1)$-dimensional Gross-Neveu model, followed by refs.~\cite{Yamamoto:2014lia,Pannullo:2019bfn}.)

The plan of the paper is as follows. In section~\ref{sec:EFT}, we introduce the class of (pseudo)real QCD-like theories and construct the corresponding low-energy EFT at the leading order of the derivative expansion. The anomalous Wess-Zumino (WZ) term, which is crucial for the existence of the CSL phase, is discussed to some detail in section~\ref{subsec:WZ}. The ground state and excitation spectrum of the EFT is then studied in the subsequent sections, first in section~\ref{sec:chiral} in the limit of vanishing quark mass (the chiral limit), and then in section~\ref{sec:away} for massive quarks. A discussion of semi-numerical variational minimization of the Hamiltonian is relegated to appendix~\ref{app:minimization}.

The main result regarding the structure of the phase diagram was already reported in ref.~\cite{Brauner:2019rjg}. In this sequel, we provide additional calculational details, in particular the derivation of the WZ term in section~\ref{subsec:WZ} and the variational minimization of the Hamiltonian in appendix~\ref{app:minimization}. The analysis of the excitation spectrum in sections~\ref{sec:chiral} and~\ref{sec:away} is completely new, and provides further supportive evidence that the phase diagram reported in ref.~\cite{Brauner:2019rjg} is correct, as well as an alternative signature of the presence of an inhomogeneous phase in the phase diagram.


\section{Effective field theory setup}
\label{sec:EFT}

In this paper, we consider a class of QCD-like theories where quarks transform in a real or pseudoreal representation of the gauge group. Examples of the class of real theories include theories with an arbitrary non-Abelian compact symmetry group and quarks in its adjoint representation, or theories based on the $\gr{G}_2$ gauge group~\cite{Holland:2003jy}. A prominent example of the class of pseudoreal theories is the so-called two-color QCD, that is a theory based on the $\gr{SU}(2)$ gauge group with quarks in the fundamental representation. We need not specify a concrete gauge group or its concrete representation though; the only assumption that we make is that the numbers of gluon and quark degrees of freedom are balanced in such a way that at zero temperature and density, the theory is in a confining phase where (approximate) chiral symmetry is spontaneously broken by the formation of a chiral condensate just like in QCD. As a consequence, the low-energy physics of the theory is dominated by the NG bosons of the global flavor symmetry.

It is well-known that for $N$ flavors of massless quarks, (pseudo)real QCD-like theories possess an enlarged global symmetry, $\mathcal G=\gr{SU}(2N)$. This includes the usual chiral symmetry of QCD, $\gr{SU}(N)_\text{L}\times\gr{SU}(N)_\text{R}\times\gr{U}(1)_B$, as a subgroup. The additional symmetry transformations, only present in (pseudo)real theories, do not commute with the baryon number generator $B$, that is, convert quarks into antiquarks and vice versa. In the ground state, the symmetry is broken to a subgroup $\mathcal H\subset\mathcal G$ by the formation of a chiral condensate. In real theories, $\mathcal H=\gr{SO}(2N)$, leading to $2N^2+N-1$ NG bosons. In pseudoreal theories, on the other hand, $\mathcal H=\gr{Sp}(2N)$, leading to $2N^2-N-1$ NG bosons. Out of these, $N^2-1$ are the usual pseudoscalar mesons, already present in QCD. The remaining NG modes correspond to scalar diquarks and antidiquarks, altogether $N(N+1)/2$ pairs in real theories and $N(N-1)/2$ ones in pseudoreal theories~\cite{Kogut:2000ek}. 

For the sake of an easy comparison with the phase structure of QCD in strong magnetic fields, we will restrict to (pseudo)real theories with $N=2$ degenerate light quark flavors. (No further restrictions on the color degrees of freedom beyond those introduced above will be imposed.) It was shown in ref.~\cite{Brauner:2019rjg} that such theories are free of the sign problem in the simultaneous presence of a baryon number chemical potential $\mu_B$ and an external magnetic field $\vek H$, provided that the electric charges of the $u$-type and $d$-type quarks satisfy the condition $q_u=-q_d$. Namely, the Euclidean Dirac operators $\mathcal D_{u,d}$ of the two quark flavors are then related by the antiunitary mapping
\begin{equation}
\mathcal D_d=(KC\gamma_5\mathcal P)\mathcal D_u(KC\gamma_5\mathcal P)^{-1},
\label{DuDd}
\end{equation}
where $C$ is the charge conjugation matrix, $K$ the operator of complex conjugation, and $\mathcal P$ a matrix that realizes the similarity transformation between the generators of the gauge group and their complex conjugates. Eq.~\eqref{DuDd} implies that $\det\mathcal D_d=(\det\mathcal D_u)^*$, and thus ensures that the determinant of the full Dirac operator of the two-flavor theory is real and non-negative. The condition $q_u=-q_d$ will be implicitly assumed in the following, unless explicitly stated otherwise.

In general, the low-energy EFT for the NG degrees of freedom is constructed as a nonlinear sigma model on the coset space $\mathcal G/\mathcal H$. It would therefore appear from our above discussion that we need two different EFTs, one for real and one for pseudoreal theories. However, our task is dramatically simplified by recalling that we are interested in the possible presence of a CSL-like phase in the phase diagram, which requires a strong background magnetic field~\cite{Brauner:2016pko}. In a sufficiently strong magnetic field, electrically charged degrees of freedom become heavy due to Landau level quantization. The low-energy physics will then be dominated by the electrically neutral NG modes. To construct their EFT description, it is sufficient to consider the subgroups of $\mathcal G$ and $\mathcal H$, left intact by the magnetic field. For two quark flavors and $q_u=-q_d\neq0$, these are the same for real and pseudoreal theories,
\begin{equation}
\mathcal G_Q=\gr{SU}(2)\times\gr{SU}(2)\times\gr{U}(1)_Q,\qquad
\mathcal H_Q=\gr{SU}(2)_\text{diag}\times\gr{U}(1)_Q,
\label{GQHG}
\end{equation}
where $\gr{U}(1)_Q$ refers to the subgroup, generated by the operator of electric charge $Q$. The low-energy spectrum contains accordingly $\dim\mathcal G_Q/\mathcal H_Q=3$ states, corresponding to the neutral pion and an electrically neutral quark-antidiquark pair. Note that the $\gr{SU(2)}\times\gr{SU(2)}$ subgroup of $\mathcal G_Q$ is \emph{not} the usual chiral symmetry of two-flavor QCD: it contains, among others, the operator of baryon number $B$ as one of its ``diagonal'' generators.

To complete the setup for the construction of the low-energy EFT, we need to know the appropriate spacetime symmetry. In strong magnetic fields, assumed here, the Lorentz group $\gr{SO(3,1)}$ is broken explicitly down to its $\gr{SO(1,1)}\times\gr{SO(2)}$ subgroup; the former factor includes boosts along the magnetic field, while the latter rotations in the transverse plane. This reduced spacetime symmetry can be implemented by constructing a Lagrangian density in the usual manner, but instead of the Minkowski metric $g_{\mu\nu}$ contracting Lorentz indices using its projections, $g_{\parallel\mu\nu}$ to the two-dimensional subspace spanning time and the direction along the magnetic field, and $g_{\perp\mu\nu}$ to the two-dimensional transverse plane.

Finally, we need a power-counting scheme to organize individual contributions to the effective Lagrangian. As usual in the chiral perturbation theory of QCD~\cite{Ecker:1994gg,Pich:1995bw,Scherer:2002tk}, we assign each derivative appearing in the Lagrangian density the order $\mathcal O(p^1)$. Likewise, the baryon chemical potential $\mu_B$ counts as $\mathcal O(p^1)$. In contrast to the chiral perturbation theory, however, we will count the external magnetic field as $\mathcal O(p^0)$. This is necessary in order to keep consistency of the EFT in strong magnetic fields. Since the magnetic field is a singlet of both the internal symmetry $\mathcal G_Q$ and the reduced spacetime symmetry, it is implemented by making the couplings in the effective Lagrangian arbitrary functions of $\vek H$.

Having put together all the necessary pieces, we can now write down the effective Lagrangian at the leading, second order of the derivative expansion~\cite{Miransky:2002rp,Miransky:2015ava},
\begin{equation}
\La_\text{eff}=\frac{f_\pi^2}4\bigl[(g^{\mu\nu}_\parallel+v^2g^{\mu\nu}_\perp)\tr(D_\mu\Sigma D_\nu\Sigma^{-1})+m_\pi^2\tr(\Sigma+\Sigma^{-1})\bigr]+\La_\text{WZ}.
\label{lagrangian}
\end{equation}
Here $\Sigma$ is the unitary unimodular $2\times2$ matrix field that includes the three degrees of freedom of the $\mathcal G_Q/\mathcal H_Q$ coset space. Its covariant derivative defines the coupling of diquarks to the baryon number chemical potential,
\begin{equation}
D_\mu\Sigma\equiv\de_\mu\Sigma-\imag\delta_{\mu0}b\mu_B[\tau_3,\Sigma],
\end{equation}
where $b$ is the baryon number of a single quark and $\tau_3$ the third Pauli matrix. The effective couplings in the Lagrangian~\eqref{lagrangian}---the pion decay constant $f_\pi$, pion mass $m_\pi$ and the velocity parameter $v$---are given by a priori unknown functions of the external magnetic field. Finally, the $\La_\text{WZ}$ term represents the contribution of the chiral anomaly, known as the Wess-Zumino (WZ) term~\cite{Wess:1971yu,Witten:1983tw}. This has to be specified concretely before we can proceed with the analysis of the EFT.


\subsection{Wess-Zumino term}
\label{subsec:WZ}

In principle we could avoid having to construct the WZ term from scratch by using the results of ref.~\cite{Son:2007ny} and merely swapping the roles of baryon number $B$ and electric charge $Q$ therein. It is, however, instructive to outline at least the main steps. From the perspective of the low-energy EFT, we can at first forget about the anomaly in the underlying microscopic theory and simply ask the question whether there are nontrivial contributions to the effective Lagrangian that, unlike the first term in eq.~\eqref{lagrangian}, are invariant under the internal symmetry group $\mathcal G_Q$ only up to a surface term~\cite{DHoker:1994ti,DHoker:1995it}.

It turns out that in four spacetime dimensions and for the internal symmetry~\eqref{GQHG}, there is a single nontrivial WZ term, determined uniquely up to the addition of an arbitrary strictly invariant operator. It can be most conveniently expressed as a differential 4-form, $\omega_\text{WZ}=A^Q\wedge\omega_\text{GW}$, where $A^Q$ is the gauge field 1-form of the $\gr{U(1)}_Q$ factor of the symmetry group and $\omega_\text{GW}$ is an invariant 3-form, corresponding to the so-called Goldstone-Wilczek current~\cite{Goldstone:1981kk}. This is given explicitly for instance in eq.~(4.5) of ref.~\cite{Brauner:2018zwr},
\begin{equation}
\omega_\text{GW}\propto\tr\bigl(-\tfrac13D\Sigma\wedge D\Sigma^{-1}\wedge D\Sigma\Sigma^{-1}+\imag D\Sigma\Sigma^{-1}\wedge F_\text{L}-\imag D\Sigma^{-1}\Sigma\wedge F_\text{R}\bigr),
\end{equation}
where $F_\text{L,R}$ are the field strength 2-forms associated with the left and right $\gr{SU(2)}$ factors of $\mathcal G_Q$, and the covariant derivative $D\Sigma$ encodes the corresponding gauge field 1-forms,
\begin{equation}
D\Sigma=\dd\Sigma-\imag A_\text{L}\Sigma+\imag\Sigma A_\text{R}.
\end{equation}
For our purposes, it is sufficient to consider gauge fields associated with the electric charge $Q$ and baryon number $B$, $A^Q$ and $A^B$, whence $A_\text{L}=A_\text{R}=b\tau_3A^B$ and $F_\text{L}=F_\text{R}=b\tau_3\dd A^B$. A straightforward manipulation then leads to an expression for the WZ Lagrangian,
\begin{equation}
\La_\text{WZ}=-\frac C6\epsilon^{\mu\nu\alpha\beta}A^Q_\mu\tr(\de_\nu\Sigma\de_\alpha\Sigma^{-1}\de_\beta\Sigma\Sigma^{-1})+\frac{\imag bC}4\epsilon^{\mu\nu\alpha\beta}F^Q_{\mu\nu}A^B_\alpha\tr[\tau_3(\de_\beta\Sigma\Sigma^{-1}-\de_\beta\Sigma^{-1}\Sigma)],
\label{WZterm}
\end{equation}
where $C$ is an as yet undetermined normalization factor.

The latter can be fixed by considering a single operator contributing to the WZ term~\eqref{WZterm} and matching it to the underlying gauge theory. A minor modification of the standard derivation of the Abelian anomaly (see section 22.2 of ref.~\cite{Weinberg:1996v2}) gives a contribution to the effective Lagrangian, linear in the neutral pion field $\pi^0$, as
\begin{equation}
\La_\text{WZ}\supset-\frac{d}{32\pi^2}\frac{\pi^0}{f_\pi}\epsilon^{\mu\nu\alpha\beta}\tr\bigl(\tau_3F_{\mu\nu}F_{\alpha\beta}\bigr),
\end{equation}
where $d$ is the dimension of the representation of the color gauge group that a single quark transforms in, and the remaining trace is performed over the space of quark flavors. The field strength $F_{\mu\nu}$ is diagonal in flavor and includes the contributions of both the electric charge and the baryon number. For future reference, we also include a contribution from an external gauge field coupled to isospin, that is, write schematically $F_{\mu\nu}=QF^Q_{\mu\nu}+BF^B_{\mu\nu}+IF^I_{\mu\nu}$. Inserting finally the proper values of the quantum numbers, $Q=\text{diag}(q_u,q_d)$, $B=\text{diag}(b,b)$ and $I=\text{diag}(1/2,-1/2)$, leads to the result,
\begin{equation}
\La_\text{WZ}\supset-\frac{d}{32\pi^2}\frac{\pi^0}{f_\pi}\epsilon^{\mu\nu\alpha\beta}\bigl[(q_u+q_d)F^Q_{\mu\nu}+2bF^B_{\mu\nu}\bigr]\bigl[(q_u-q_d)F^Q_{\alpha\beta}+F^I_{\alpha\beta}\bigr].
\label{WZterm2}
\end{equation}
This result applies to all QCD-like theories with two quark flavors regardless of the color gauge group; setting $d=3$, $b=1/3$, $q_u=2/3$ and $q_d=-1/3$ reproduces the known result for QCD itself~\cite{Son:2007ny}. To match it to the form of WZ term~\eqref{WZterm} valid for (pseudo)real QCD-like theories, we insert $\Sigma=\exp\bigl(\frac\imag{f_\pi}\tau_3\pi^0\bigr)$ therein and integrate by parts. For $q_u=-q_d$ as assumed in this paper, the purely electromagnetic operator $\pi^0\epsilon^{\mu\nu\alpha\beta}F^Q_{\mu\nu}F^Q_{\alpha\beta}$, responsible for the two-photon decay of neutral pions, is missing. Comparing the coefficients of the operator $\epsilon^{\mu\nu\alpha\beta}F^Q_{\mu\nu}A^B_\alpha\de_\beta\pi^0$ in eqs.~\eqref{WZterm} and~\eqref{WZterm2} then gives
\begin{equation}
C=\frac{d}{8\pi^2}(q_u-q_d).
\label{Cdef}
\end{equation}
The construction of our EFT is completed by replacing $A^B_\mu$ with $(\mu_B,\vek0)$. Note that due to the modified power counting for the background magnetic field, the WZ term~\eqref{WZterm} indeed belongs to the leading order Lagrangian~\eqref{lagrangian} as advertised.


\subsection{Effective Lagrangian and Hamiltonian}
\label{subsec:efflag}

The remainder of the paper is devoted to an analysis of the EFT as defined by eqs.~\eqref{lagrangian}, \eqref{WZterm} and \eqref{Cdef}. We will from now on assume without loss of generality that $q_u=-q_d=1/2$ and $b=1/2$. This convention is natural in that it maintains the physical electric charges of the three pions, and gives the diquarks a unit baryon number. Any other choice can be absorbed into a redefinition of $\vek H$ and $\mu_B$, respectively.

To prepare for the analysis, we first rewrite the Lagrangian~\eqref{lagrangian} in terms of a unit vector variable $(n_0,\vec n)$, defined by
\begin{equation}
\Sigma=n_0+\imag\vec n\cdot\vec\tau,\qquad
n_0^2+\vec n^2=1.
\end{equation}
This brings the Lagrangian to the form
\begin{equation}
\begin{split}
\label{Leff}
\La_\text{eff}={}&\frac{f_\pi^2}2(g^{\mu\nu}_\parallel+v^2g^{\mu\nu}_\perp)(\de_\mu n_0\de_\nu n_0+\de_\mu\vec n\cdot\de_\nu\vec n)+f_\pi^2\mu_B(n_1\de_0n_2-n_2\de_0n_1)\\
&+\frac{f_\pi^2}2\mu_B^2(n_1^2+n_2^2)+f_\pi^2m_\pi^2(n_0-1)+CH\mu_B(n_0\de_zn_3-n_3\de_zn_0)+\dotsb,
\end{split}
\end{equation}
where we oriented the magnetic field along the $z$-axis, and subtracted a constant to ensure that the Lagrangian vanishes in the vacuum where $(n_0,\vec n)=(1,\vec 0)$. Finally, the ellipsis stands for three-derivative contributions coming from the WZ term~\eqref{WZterm}. We will drop these contributions, as they have no effect on either the structure of the ground state or the excitation spectrum. Indeed, choosing a gauge where $A^Q_0=0$, it is easy to see that they are necessarily linear in time derivatives, and thus do not contribute to the Hamiltonian. Moreover, we will show below that the ground state of the EFT is either uniform or modulated in the direction of the magnetic field. When analyzing the excitation spectrum, each of the three derivatives in the three-derivative terms then either acts on the $z$-dependence of the ground state, or generates a factor of momentum; their contributions then vanish by antisymmetry.

For the record, we also write down another form of the Lagrangian, where the unit vector $(n_0,\vec n)$ is mapped on three angular variables via
\begin{equation}
n_0=\cos\theta\cos\phi,\qquad
n_1=\sin\theta\cos\alpha,\qquad
n_2=\sin\theta\sin\alpha,\qquad
n_3=\cos\theta\sin\phi.
\label{angle}
\end{equation}
This converts the Lagrangian~\eqref{Leff}, now dropping the three-derivative terms, into
\begin{equation}
\begin{split}
\La_\text{eff}={}&\frac{f_\pi^2}2(g^{\mu\nu}_\parallel+v^2g^{\mu\nu}_\perp)(\de_\mu\theta\de_\nu\theta+\de_\mu\phi\de_\nu\phi\cos^2\theta+\de_\mu\alpha\de_\nu\alpha\sin^2\theta)+f_\pi^2\mu_B\de_0\alpha\sin^2\theta\\
&+\frac{f_\pi^2}2\mu_B^2\sin^2\theta+f_\pi^2m_\pi^2(\cos\theta\cos\phi-1)+CH\mu_B\de_z\phi\cos^2\theta.
\end{split}
\label{Leff2}
\end{equation}
Upon carrying out a Legendre transform, it is easy to see that derivatives with respect to time and the transverse directions only enter the Hamiltonian in squares. The ground state is thus necessarily independent of both time and the transverse coordinates, and can be found by minimization of the spatial average of the one-dimensional effective Hamiltonian,
\begin{equation}
\begin{split}
\Ha^\text{(1d)}_\text{eff}={}&\frac{f_\pi^2}2\bigl[(\de_z\theta)^2+(\de_z\phi)^2\cos^2\theta+(\de_z\alpha)^2\sin^2\theta\bigr]\\
&-\frac{f_\pi^2}2\mu_B^2\sin^2\theta+f_\pi^2m_\pi^2(1-\cos\theta\cos\phi)-CH\mu_B\de_z\phi\cos^2\theta.
\end{split}
\label{Heff}
\end{equation}


\section{Chiral limit}
\label{sec:chiral}

\subsection{Ground state}
\label{subsec:chiralground}

In the chiral limit where $m_\pi=0$, the ground state can be found straightforwardly by rewriting the Hamiltonian~\eqref{Heff} as
\begin{equation}
\begin{split}
\left.\Ha^\text{(1d)}_\text{eff}\right|_{m_\pi=0}={}&\frac{f_\pi^2}2\bigl[(\de_z\theta)^2+(\de_z\phi-\bar H\mu_B)^2\cos^2\theta+(\de_z\alpha)^2\sin^2\theta\bigr]\\
&+\frac{f_\pi^2}2\mu_B^2\bigl[-1+(1-\bar H^2)\cos^2\theta\bigr],
\end{split}
\label{Heffchiral}
\end{equation}
where we introduced the shorthand notation
\begin{equation}
\bar H\equiv\frac{CH}{f_\pi^2}.
\label{barH}
\end{equation}
From the first line of eq.~\eqref{Heffchiral}, it is obvious that the ground state is realized for constant $\theta$. At the same time, the expectation value of the angle $\phi$ should satisfy $\de_z\langle\phi\rangle=\bar H\mu_B$ as long as $\cos\langle\theta\rangle\neq0$. The second line of eq.~\eqref{Heffchiral} picks the most favorable value of $\theta$, depending on whether $\bar H$ is smaller or greater than one. Altogether, we find two different phases:
\begin{itemize}
\item $\bar H<1$: the ground state is at $\langle\theta\rangle=\pi/2$, corresponding to
\begin{equation}
\langle n_0\rangle=0,\qquad
\langle n_1\rangle=\cos\langle\alpha\rangle,\qquad
\langle n_2\rangle=\sin\langle\alpha\rangle,\qquad
\langle n_3\rangle=0
\label{BECchiral}
\end{equation}
with constant $\langle\alpha\rangle$. This describes a Bose-Einstein condensate (BEC) of diquarks.
\item $\bar H>1$: the ground state is at $\langle\theta\rangle=0$, corresponding to
\begin{equation}
\langle n_0\rangle=\cos(\bar H\mu_B z),\qquad
\langle n_1\rangle=\langle n_2\rangle=0,\qquad
\langle n_3\rangle=\sin(\bar H\mu_Bz),
\label{CSLchiral}
\end{equation}
up to a shift of the $z$-coordinate. This describes the CSL state in the chiral limit~\cite{Brauner:2016pko}.
\end{itemize}
At any nonzero value of the baryon chemical potential, the phase diagram of the EFT in the $\bar H$-$\mu_B$ plane (and at zero temperature) therefore contains two phases. The $\mu_B$-independent critical magnetic field for the formation of the inhomogeneous CSL state is, by eq.~\eqref{barH},
\begin{equation}
H_\text{crit}=\frac{8\pi^2f_\pi^2}d.
\label{Hcrit}
\end{equation}
The parametric dependence on the number of quark color degrees of freedom, encoded in the constant $d$, arises from the competition of the anomalous and non-anomalous parts of the effective Hamiltonian, only the former of which is proportional to $d$. We will return to this observation in the conclusions.


\subsection{Excitation spectrum}
\label{subsec:chiralspectrum}

The angular parameterization~\eqref{angle} becomes singular at both $\theta=0$ and $\theta=\pi/2$. In order to discuss the fluctuations about the ground state, we therefore have to return to the unit-vector parameterization, in which the Lagrangian takes the form~\eqref{Leff}.

Let us first focus on the BEC phase~\eqref{BECchiral}. By setting $\langle\alpha\rangle$ without loss of generality to zero, we can parameterize $n_1$ and $n_2$ as
\begin{equation}
n_1=\rho\cos\alpha,\qquad
n_2=\rho\sin\alpha,\qquad
\text{where }
\rho\equiv\sqrt{1-n_0^2-n_3^2}.
\end{equation}
The effective Lagrangian~\eqref{Leff} can then be expanded in powers of the independent fluctuations $n_0$, $n_3$ and $\alpha$. Keeping only terms bilinear in these fluctuations, we obtain
\begin{equation}
\begin{split}
\La_\text{bilin}^\text{BEC}={}&\frac{f_\pi^2}2(g^{\mu\nu}_\parallel+v^2g^{\mu\nu}_\perp)(\de_\mu n_0\de_\nu n_0+\de_\mu n_3\de_\nu n_3+\de_\mu\alpha\de_\nu\alpha)\\
&-\frac{f_\pi^2}2\mu_B^2(n_0^2+n_3^2)+CH\mu_B(n_0\de_zn_3-n_3\de_zn_0).
\end{split}
\end{equation}
Obviously, $\alpha$ describes a gapless pseudo-relativistic excitation with the dispersion relation
\begin{equation}
\omega_\alpha(\vek p)=\sqrt{v^2\vek p_\perp^2+p_z^2}.
\label{dispNG}
\end{equation}
This is the NG boson of the spontaneously broken $\gr{U(1)}_B$ symmetry. The $n_0$ and $n_3$ modes are mixed by the anomaly term. Diagonalizing the $2\times2$ inverse propagator leads to a pair of dispersion relations,
\begin{equation}
\omega_\pm(\vek p)=\sqrt{v^2\vek p_\perp^2+(p_z\pm\bar H\mu_B)^2+(1-\bar H^2)\mu_B^2}.
\label{disppion}
\end{equation}
Both these modes become gapless at the phase transition to the CSL phase, where $\bar H=1$. The nonuniform nature of the CSL state is manifested by the fact that the minimum of $\omega_\pm(\vek p)$ appears at a nonzero value of $p_z$, exactly corresponding to the momentum scale of the CSL solution~\eqref{CSLchiral}. Note that the energy of these excitations at zero momentum is fully determined by the baryon chemical potential alone, $\omega_\pm(\vek p=\vek0)=\mu_B$, in accord with the interpretation of both modes as massive NG bosons~\cite{Watanabe:2013uya}.

Moving on to the CSL phase, we need to find a suitable parameterization of fluctuations about the ground state~\eqref{CSLchiral}. This time, we can set
\begin{equation}
n_0=\rho\cos(\bar H\mu_Bz+\phi),\qquad
n_3=\rho\sin(\bar H\mu_Bz+\phi),\qquad
\text{where }
\rho\equiv\sqrt{1-n_1^2-n_2^2}.
\label{CSLchiralparam}
\end{equation}
The three independent excitation modes then correspond to $n_1$, $n_2$ and $\phi$. The bilinear part of the Lagrangian~\eqref{Leff} is then, upon some manipulation, seen to be
\begin{equation}
\begin{split}
\La_\text{bilin}^\text{CSL}={}&\frac{f_\pi^2}2(g^{\mu\nu}_\parallel+v^2g^{\mu\nu}_\perp)(\de_\mu n_1\de_\nu n_1+\de_\mu n_2\de_\nu n_2+\de_\mu\phi\de_\nu\phi)\\
&-\frac{f_\pi^2}2(\bar H^2-1)\mu_B^2(n_1^2+n_2^2)+f_\pi^2\mu_B(n_1\de_0n_2-n_2\de_0n_1).
\end{split}
\end{equation}
We find again one gapless pseudo-relativistic excitation with the dispersion relation
\begin{equation}
\omega_\phi(\vek p)=\sqrt{v^2\vek p_\perp^2+p_z^2}.
\end{equation}
This pion-like mode can be interpreted as the NG boson of the spontaneously broken translation invariance in the CSL state. The $n_1$ and $n_2$ modes describe the diquark-antidiquark pair, and their dispersion relation is obtained by diagonalization,
\begin{equation}
\omega_{d,\bar d}(\vek p)=\sqrt{v^2\vek p_\perp^2+p_z^2+\bar H^2\mu_B^2}\mp\mu_B.
\end{equation}
We can see that the diquark mode becomes gapless at the phase transition, $\bar H=1$, signaling an instability towards the BEC state.


\section{Away from the chiral limit}
\label{sec:away}

In order to investigate the phase diagram away from the chiral limit, we take advantage of the angular parameterization~\eqref{angle}. The analysis is simplified by introducing dimensionless spacetime coordinates $\bar x^\mu$ and a dimensionless chemical potential $x$ via\footnote{The use of the letter $x$ as a chemical potential may be somewhat unfortunate here, but nevertheless is established in the literature.}
\begin{equation}
\bar x^\mu\equiv m_\pi x^\mu,\qquad
x\equiv\frac{\mu_B}{m_\pi}.
\end{equation}
The effective one-dimensional Hamiltonian~\eqref{Heff} then also acquires a dimensionless form,
\begin{equation}
\begin{split}
\bar\Ha&\equiv\frac{\Ha_\text{eff}^\text{(1d)}}{f_\pi^2m_\pi^2}\\
&=\frac12\left[(\theta')^2+(\phi')^2\cos^2\theta+(\alpha')^2\sin^2\theta\right]-\frac{x^2}2\sin^2\theta+1-\cos\theta\cos\phi-x\bar H\phi'\cos^2\theta,
\end{split}
\label{HamDimless}
\end{equation}
where the primes denote derivatives with respect to $\bar z$.


\subsection{Ground state}
\label{subsec:awayground}

Finding the absolute minimum of (the spatial average of) $\bar\Ha$ seems difficult in a closed form, we will therefore have to make some further assumptions about the structure of the ground state. We have performed extensive checks, both numeric and where possible analytic, that no state with energy lower than the states discussed in detail below exists for any choice of the dimensionless parameters $x$ and $\bar H$. Some details of these checks are provided for an interested reader in appendix~\ref{app:minimization}.

The analysis becomes elementary if we for a moment restrict to uniform field configurations. Finding the ground state then boils down to a minimization of the static part of the Hamiltonian~\eqref{HamDimless} with respect to $\theta$ and $\phi$. We find two different stationary states:
\begin{itemize}
\item Trivial vacuum, where
\begin{equation}
\langle\theta\rangle=\langle\phi\rangle=0,\qquad
\bar\Ha_\text{vac}=0.
\end{equation}
\item BEC of diquarks, where
\begin{equation}
\cos\langle\theta\rangle=\frac1{x^2},\qquad
\langle\phi\rangle=0,\qquad
\bar\Ha_\text{BEC}=-\frac12\left(x-\frac1x\right)^2.
\end{equation}
\end{itemize}
The BEC state only exists as a stationary state of the Hamiltonian if $x>1$, or $\mu_B>m_\pi$, and it automatically has a lower energy than the trivial vacuum. This exhausts the possible candidate ground states in the subspace of uniform field configurations. To prove that, for given values of $x$ and $\bar H$, the ground state is inhomogeneous, it is therefore sufficient to find \emph{any} state that has a lower energy than the two states listed above.

\begin{figure}
\begin{center}
\includegraphics[scale=1]{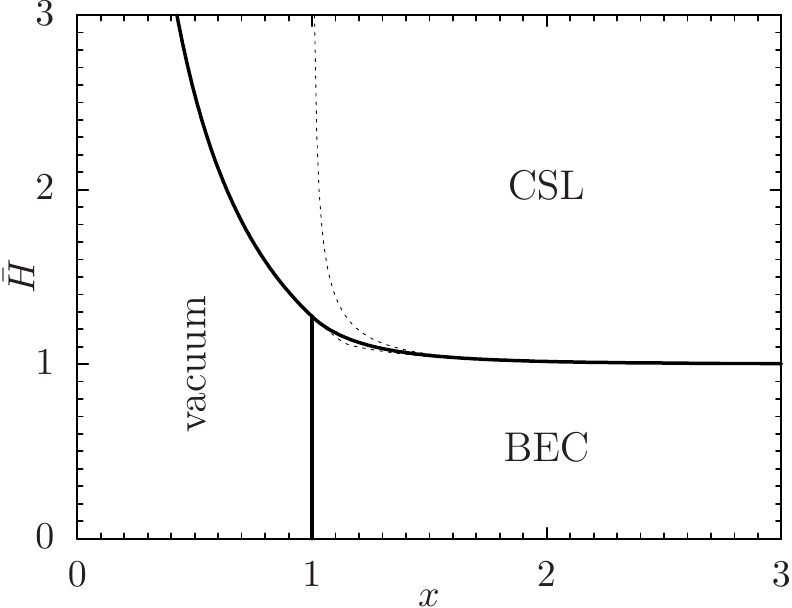}
\end{center}
\caption{Phase diagram of the EFT, and thus also of the class of (pseudo)real QCD-like theories considered in this paper, as a function of the dimensionless magnetic field and chemical potential. The figure is adapted from ref.~\cite{Brauner:2019rjg}. The solid lines indicate phase transitions. The dashed lines indicate the spinodal curves of the first-order transition between the BEC and CSL phases, derived in section~\ref{subsec:awayspectrum}.}
\label{fig:phase_diagram}
\end{figure}

To that end, we inspect a different class of states by setting $\theta\to0$. Upon this replacement, the effective Hamiltonian~\eqref{HamDimless} reduces to that of QCD where only the neutral pion degree of freedom survives in the low-energy EFT. The minimum of the Hamiltonian on such a subset of field configurations is thus known, and corresponds to the CSL state~\cite{Brauner:2016pko},
\begin{equation}
\langle\theta\rangle=0,\qquad
\cos\frac{\langle\phi(\bar z)\rangle}2=\sn\left(\frac{\bar z}k,k\right),
\label{CSLsolution}
\end{equation}
where $\sn$ is one of Jacobi's elliptic functions and $k$ the associated elliptic modulus. This is fixed by a minimization of the spatially averaged Hamiltonian, which leads to the condition,
\begin{equation}
\frac{E(k)}k=\frac{\pi x\bar H}4,
\label{CSLvacuum}
\end{equation}
where $E(k)$ is the complete elliptic integral of the second kind. Using the explicit solution~\eqref{CSLsolution} along with eq.~\eqref{CSLvacuum}, the spatially averaged energy density of the CSL state can be cast as
\begin{equation}
\bar\Ha_\text{CSL}=2\left(1-\frac1{k^2}\right).
\end{equation}
Given that the elliptic modulus falls into the range $0\leq k\leq1$, this state always has a lower energy than the trivial vacuum. Since $E(k)/k\geq1$, it only exists for $\bar H\geq4/(\pi x)$ though.

Comparing the energies of all the three candidate states gives rise to the phase structure shown in figure~\ref{fig:phase_diagram}. We can see that there is a range of magnetic fields where, as anticipated, the ground state becomes inhomogeneous. The critical magnetic field for the formation of an inhomogeneous ground state is given by
\begin{equation}
\begin{split}
x&<1:\qquad
\bar H_\text{crit}=\frac4{\pi x},\\
x&>1:\qquad
\bar H_\text{crit}=\frac4{\pi x}\frac{E(k_0)}{k_0},\quad
\text{where }
k_0=\frac2{x+\frac1x}.
\end{split}
\end{equation}
Note that the limit $x\to\infty$ is equivalent to the chiral limit, $m_\pi\to0$. For large $x$, the critical magnetic field separating the BEC and CSL phases therefore goes to $\bar H=1$.

\begin{figure}
\begin{center}
\includegraphics[scale=1]{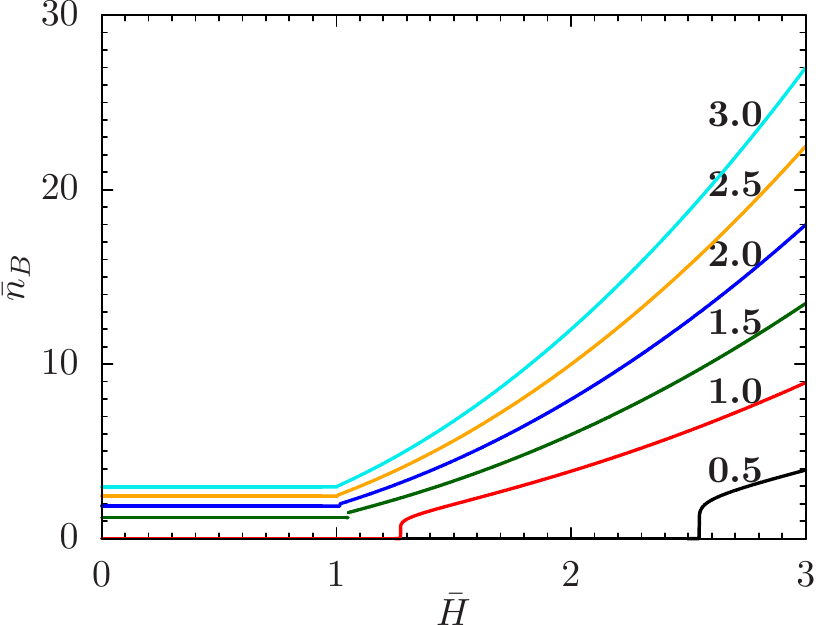}
\end{center}
\caption{Spatially averaged baryon number density in the ground state as a function of magnetic field for several different values of the dimensionless chemical potential $x$ (indicated in bold). The first-order phase transition from the BEC to the CSL phase for $x>1$ is visible in the curves.}
\label{fig:nB}
\end{figure}

Except for the vacuum phase, the ground state is characterized by nonzero baryon number density, which can be obtained from the Hamiltonian~\eqref{Heff} by taking a derivative with respect to the chemical potential. Defining a rescaled, dimensionless baryon density as $\bar n_B\equiv n_B/(f_\pi^2m_\pi)$, its value in the BEC phase, and the spatially averaged value in the CSL phase~\cite{Brauner:2016pko}, can be expressed as\footnote{The expression for $\bar n_B^\text{BEC}$ was first obtained in ref.~\cite{Kogut:2000ek}, see eq.~(107) therein.}
\begin{equation}
\bar n_B^\text{BEC}=x\left(1-\frac1{x^4}\right),\qquad
\bar n_B^\text{CSL}=\frac{\pi\bar H}{kK(k)},
\end{equation}
where $K(k)$ is the complete elliptic integral of the first kind, and $k$ is determined implicitly by eq.~\eqref{CSLvacuum}. The dependence of the baryon number density on magnetic field for several different values of chemical potential is shown in figure~\ref{fig:nB}. The numerical results show that while in the BEC phase, the baryon density is driven solely by the chemical potential, in the CSL phase, it is further boosted by increasing magnetic field. This is in accord with eq.~\eqref{Heffchiral} which implies that in the chiral limit, $n_B=f_\pi^2\mu_B\bar H^2$. The corresponding dimensionless density, $\bar n_B=x\bar H^2$, describes accurately the curves shown in figure~\ref{fig:nB}.


\subsection{Excitation spectrum}
\label{subsec:awayspectrum}

It is of course possible that by restricting to the three candidate states as discussed above, we miss a phase in the phase diagram where a more sophisticated, necessarily inhomogeneous, ground state has an even lower energy. Apart from the direct checks reported in appendix~\ref{app:minimization}, one can also check independently the consistency of the phase diagram displayed in figure~\ref{fig:phase_diagram} by evaluating the spectrum of excitations above the tentative ground state. Should this reveal an instability in some of the dispersion branches, we would have a solid evidence for the existence of a new, more favorable ground state.

Let us start with the region in the phase diagram where the trivial vacuum seems to prevail. This corresponds to $\langle n_0\rangle=1$ and $\langle\vec n\rangle=\vec0$. To extract the spectrum, we return to eq.~\eqref{Leff} and use $\vec n$ as the three independent degrees of freedom along with $n_0=\sqrt{1-\vec n^2}$. The bilinear part of the effective Lagrangian~\eqref{Leff} then reads
\begin{equation}
\La^\text{vac}_\text{bilin}=f_\pi^2\biggl[\frac12(g^{\mu\nu}_\parallel+v^2g^{\mu\nu}_\perp)\de_\mu\vec n\cdot\de_\nu\vec n+\mu_B(n_1\de_0 n_2-n_2\de_0 n_1)+\frac12\mu_B^2(n_1^2+n_2^2)-\frac12m_\pi^2\vec n^2\biggr].
\end{equation}
This is a known result: the spectrum contains a neutral pion with mass $m_\pi$ and a diquark-antidiquark pair with mass $m_\pi$ and chemical potential $\mu_B$. We can readily identify the instability towards diquark BEC at $\mu_B=m_\pi$. On the contrary, there is nothing in the spectrum that would indicate the onset of the inhomogeneous CSL phase.


\subsubsection{Spectrum in the BEC phase}

The spectrum of the BEC phase can be analyzed in the same manner, except that we can now take advantage of the angular parameterization~\eqref{angle}. Denoting the value of $\theta$ in the BEC ground state as $\theta_0$ and expanding in the fluctuations of all the three angles $\theta$, $\phi$, $\alpha$, a straightforward manipulation based on eq.~\eqref{Leff2} gives the rescaled bilinear Lagrangian
\begin{equation}
\begin{split}
\bar\La^\text{BEC}_\text{bilin}\equiv\frac{\La^\text{BEC}_\text{bilin}}{f_\pi^2m_\pi^2}={}&\frac12(g^{\mu\nu}_\parallel+v^2g^{\mu\nu}_\perp)\bar\de_\mu\theta\bar\de_\nu\theta-\frac12x^2\theta^2\sin^2\theta_0\\
&+\frac12(g^{\mu\nu}_\parallel+v^2g^{\mu\nu}_\perp)\bar\de_\mu\phi\bar\de_\nu\phi\cos^2\theta_0-\frac12\phi^2\cos\theta_0\\
&+\frac12(g^{\mu\nu}_\parallel+v^2g^{\mu\nu}_\perp)\bar\de_\mu\alpha\bar\de_\nu\alpha\sin^2\theta_0+x\sin2\theta_0(\theta\bar\de_0\alpha-\bar H\theta\bar\de_z\phi).
\end{split}
\end{equation}
This readily leads to the inverse matrix propagator in the $(\theta,\phi,\alpha)$ space,
\begin{equation}
\Ge^{-1}=\begin{pmatrix}
-\bar\Box_v-x^2\sin^2\theta_0 & -x\bar H\sin2\theta_0\bar\de_z & +x\sin2\theta_0\bar\de_0\\
+x\bar H\sin2\theta_0\bar\de_z & -\bar\Box_v\cos^2\theta_0-\cos\theta_0 & 0\\
-x\sin2\theta_0\bar\de_0 & 0 & -\bar\Box_v\sin^2\theta_0
\end{pmatrix},
\end{equation}
where we used the shorthand notation $\bar\Box_v\equiv(g^{\mu\nu}_\parallel+v^2g^{\mu\nu}_\perp)\bar\de_\mu\bar\de_\nu$. Upon Fourier transforming to the space of frequency $\bar\omega$ and momentum $\bar{\vek p}$, this becomes
\begin{equation}
\Ge^{-1}(\bar\omega,\bar{\vek p})=\begin{pmatrix}
\bar\omega^2-\bar{\vek p}_v^2-x^2\sin^2\theta_0 & -\imag x\bar H\bar p_z\sin2\theta_0 & -\imag x\bar\omega\sin2\theta_0\\
+\imag x\bar H\bar p_z\sin2\theta_0 & (\bar\omega^2-\bar{\vek p}_v^2)\cos^2\theta_0-\cos\theta_0 & 0\\
+\imag x\bar\omega\sin2\theta_0 & 0 & (\bar\omega^2-\bar{\vek p}_v^2)\sin^2\theta_0
\end{pmatrix},
\label{invprop}
\end{equation}
where we likewise abbreviated $\bar{\vek p}_v^2\equiv v^2\bar{\vek p}_\perp^2+\bar p_z^2$.

The dispersion relations of the three excitation branches in dimensionless form can in principle be obtained by setting the determinant of the inverse propagator in eq.~\eqref{invprop} to zero. It is unfortunately not possible to write the result, corresponding to the solution of a cubic equation for $\bar\omega^2$, in a closed real form.\footnote{A closed analytic solution exists in the special cases of motion in the transverse plane ($\bar p_z=0$) or zero magnetic field ($\bar H=0$). In this case, the inverse propagator~\eqref{invprop} becomes block-diagonal and the ensuing dispersion relations are in accord with the results of ref.~\cite{Kogut:2000ek}, see eq.~(86) therein.} However, what interests us most is whether some of the modes becomes gapless in a part of the tentative BEC phase. To search for such an instability, it is sufficient to set $\bar\omega=0$, upon which the determinant of the inverse propagator is easily evaluated and factorized,
\begin{equation}
\det\Ge^{-1}(0,\bar{\vek p})=-\bar{\vek p}_v^2\bigl[(\bar{\vek p}_v^2+x^2)(\bar{\vek p}_v^2+x^2\sin^2\theta_0)-4x^2\bar H^2\bar p_z^2\sin^2\theta_0\bigr]\sin^2\theta_0\cos^2\theta_0.
\end{equation}
The leading factor of $\bar{\vek p}_v^2$ reflects the presence of the gapless mode due to the spontaneously broken exact $\gr{U(1)}_B$ symmetry. The expression in the square brackets is manifestly positive for $\bar H=0$, we are thus looking for the lowest value of $\bar H$, $\bar H_\text{inst}$, such that
\begin{equation}
(\bar{\vek p}_v^2+x^2)(\bar{\vek p}_v^2+x^2\sin^2\theta_0)-4x^2\bar H^2\bar p_z^2\sin^2\theta_0
\end{equation}
drops to zero for some $\bar{\vek p}$. A straightforward function analysis leads to the result,
\begin{equation}
\bar H_\text{inst}=\frac{1+\sin\theta_0}{2\sin\theta_0}.
\end{equation}
The value of $\bar H_\text{inst}$ is indicated by the upper dashed line in figure~\ref{fig:phase_diagram}. This confirms that the diquark BEC state is stable under small field fluctuations everywhere in the tentative BEC phase. The critical field $\bar H_\text{inst}$ merely denotes a point where the BEC state ceases to be a local minimum of the Hamiltonian inside the CSL phase; this defines the spinodal curve for the first-order phase transition between the BEC and CSL phases.

\begin{figure}
\begin{center}
\includegraphics[scale=1]{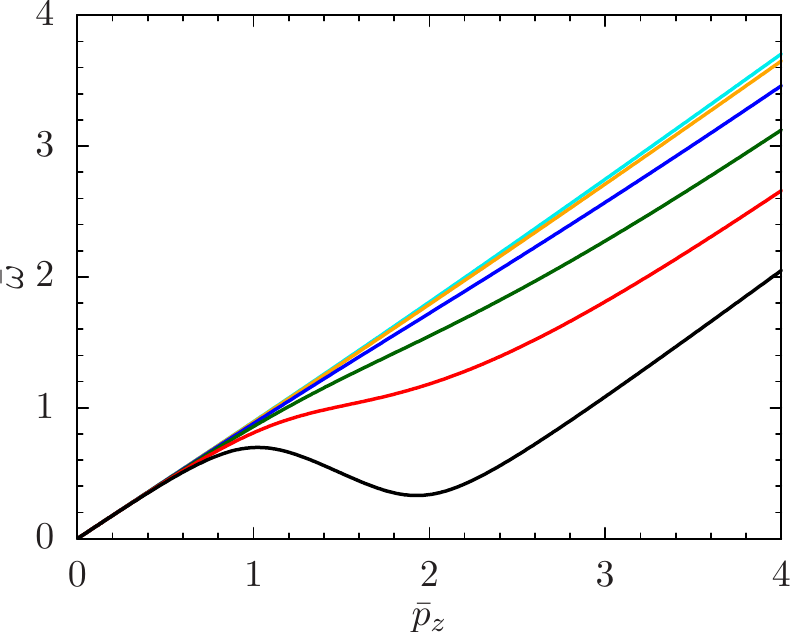}
\end{center}
\caption{Dispersion relation of the lightest excitation branch in the BEC phase at $\vek p_\perp=\vek0$ and $x=2$ for $\bar H=0$ (cyan), $0.2$ (orange), $0.4$ (blue), $0.6$ (green), $0.8$ (red), and $1$ (black).}
\label{fig:NGdisp}
\end{figure}

In order to illustrate the onset of instability as the magnetic field approaches $\bar H_\text{inst}$, we show in figure~\ref{fig:NGdisp} the evolution of the dispersion relation of the lightest mode in the spectrum, obtained by numerically solving the condition that the determinant of the inverse propagator~\eqref{invprop} is zero. We can see that close to the critical magnetic field, the dispersion relation develops a characteristic roton-like minimum, which acts as a precursor to the instability with respect to condensation of modes with nonzero momentum.

\begin{figure}
\begin{center}
\includegraphics[width=\textwidth]{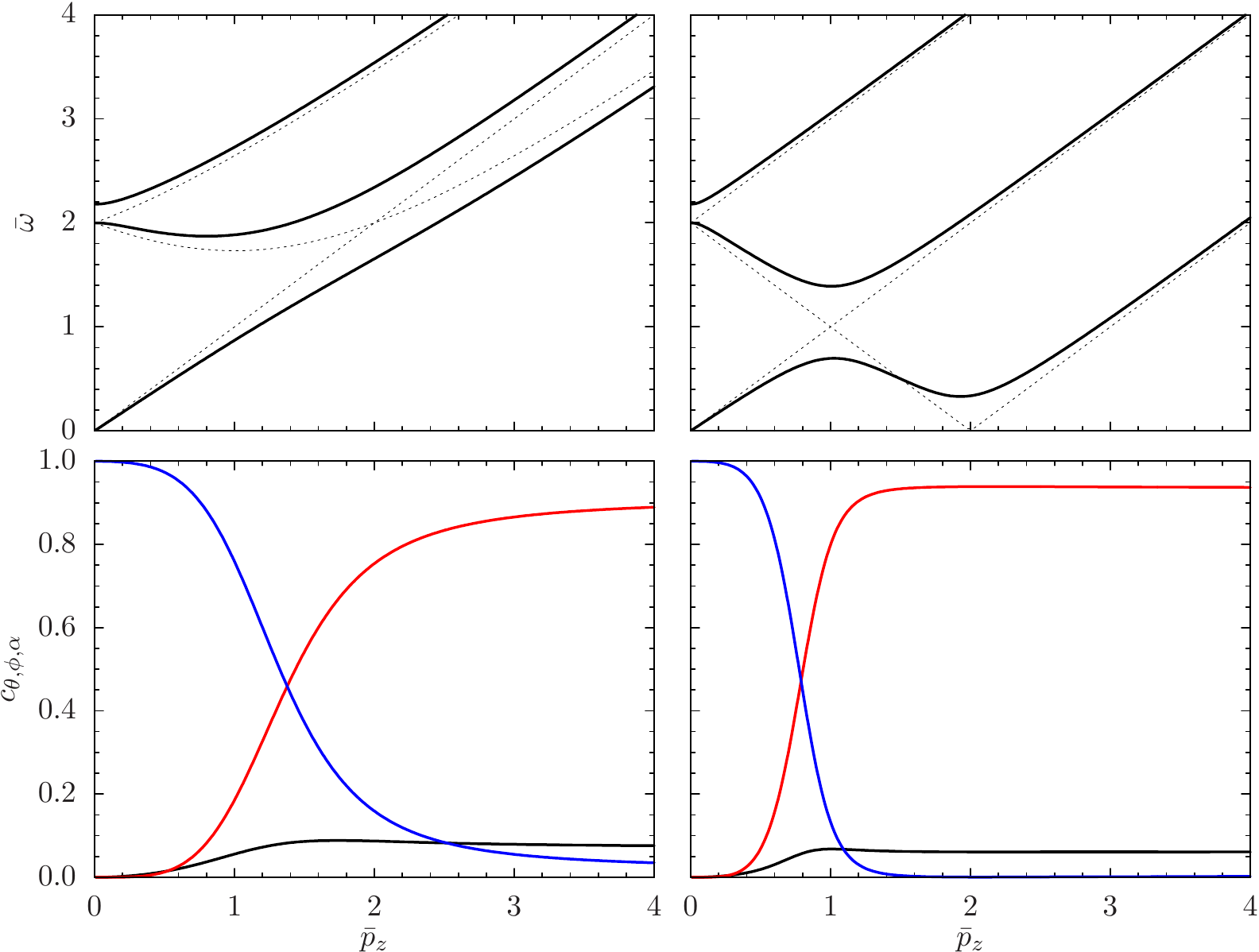}
\end{center}
\caption{Dispersion relations of the three light modes and the composition of the lightest mode for $x=2$ and $\bar H=0.5$ (left panels) or $\bar H=1$ (right panels). Upper panels: the dispersion relations obtained from the inverse propagator~\eqref{invprop} are indicated by the thick solid lines, whereas the dashed lines show for comparison the results in the chiral limit, given by eqs.~\eqref{dispNG} and~\eqref{disppion}. Lower panels: values of the coefficients $c_{\theta,\phi,\alpha}$, obtained from the spectral representation~\eqref{kallen}; $c_\theta$ is shown in black, $c_\phi$ in red and $c_\alpha$ in blue.}
\label{fig:dispersions}
\end{figure}

A cautious reader might be puzzled by the fact that the instability at $\bar H_\text{inst}$ occurs in the dispersion relation of the lightest mode, which corresponds to the NG boson of the spontaneously broken exact $\gr{U(1)}_B$ symmetry. Namely, our previous analysis of the chiral limit in section~\ref{subsec:chiralspectrum} clearly showed that the instability appears in the dispersion relation of modes that have a nonzero gap at $\vek p=\vek 0$, not of the NG mode, see eqs.~\eqref{dispNG} and~\eqref{disppion}. The resolution of this apparent contradiction lies in the mixing of all three modes, present away from the chiral limit. This is illustrated by the upper two panels of figure~\ref{fig:dispersions}, where we display the dispersion relations of all three light modes for fixed $x$ and two different values of $\bar H$. The results in the chiral limit are indicated by the dashed lines.\footnote{This is a slight abuse of notation, as the variables $\bar\omega$ and $\bar p_z$ are defined by rescaling by $m_\pi$, which seems to make little sense in the chiral limit. However, a glance at eqs.~\eqref{dispNG} and~\eqref{disppion} shows that in the chiral limit, the dispersion relations are homogeneous functions of degree one in the variables $\vek p$ and $\mu_B$. In the chiral limit, we can therefore think of $\bar\omega$ and $\bar p_z$ as being defined up to an arbitrary common scale.} We can see that switching on a nonzero quark mass leads to avoided level crossing in the spectrum, as a result of which the roton-like minimum indeed appears in the dispersion relation of the lightest mode.

To get further insight into the nature of the mixing, we extracted the coupling of the lightest mode to the $\theta$, $\phi$ and $\alpha$ fields using the K\"all\'en-Lehmann spectral representation. Schematically, the inverse of eq.~\eqref{invprop} near a pole $\bar\omega=\epsilon(\bar{\vek p})$ takes the form
\begin{equation}
\Ge_{ij}(\bar\omega,\bar{\vek p})\propto\frac{\langle0|\chi_i|\bar{\vek p}\rangle\langle\bar{\vek p}|\chi_j|0\rangle}{\bar\omega-\epsilon(\bar{\vek p})},
\label{kallen}
\end{equation}
where $\chi_i$ runs over $\theta,\phi,\alpha$. In the lower two panels of figure~\ref{fig:dispersions}, we show the values of $c_i\propto|\langle0|\chi_i|\bar{\vek p}\rangle|^2$, normalized so that $c_\theta+c_\phi+c_\alpha=1$. With this definition, the displayed coefficients highlight the composition of the lightest mode in the spectrum. Obviously, it is ``mostly $\alpha$'' before the avoided crossing (in accordance with the NG nature of the mode), and ``mostly $\phi$'' after the avoided crossing.


\subsubsection{Spectrum in the CSL phase}

The spectrum in the CSL phase can be analyzed analogously, except that we now have to deal with the fact that we are looking for fluctuations of an inhomogeneous ground state. Given that $\langle\theta\rangle=0$ in the CSL state~\eqref{CSLsolution}, the parameterization~\eqref{angle} is singular and we have to return once more to the unit vector $(n_0,\vec n)$. Denoting the spatially-dependent expectation value of $\phi$ in the CSL ground state as $\phi_0$, we can use a parameterization that straightforwardly generalizes eq.~\eqref{CSLchiralparam},
\begin{equation}
n_0=\rho\cos(\phi_0+\phi),\qquad
n_3=\rho\sin(\phi_0+\phi),\qquad
\text{where }
\rho\equiv\sqrt{1-n_1^2-n_2^2}.
\end{equation}
The three independent degrees of freedom can be taken as $n_1$, $n_2$ and $\phi$. However, since the exact baryon number $\gr{U(1)}_B$ symmetry remains unbroken in the CSL state, it is more convenient to trade $n_1$, $n_2$ for the baryon number eigenstates,
\begin{equation}
n_\pm\equiv\frac1{\sqrt2}(n_1\pm\imag n_2).
\end{equation}
The bilinear part of the effective Lagrangian then splits into separate pieces for $\phi$ and $n_\pm$, $\La^\text{CSL}_\text{bilin}=\La^\text{CSL}_{\phi}+\La^\text{CSL}_{n_\pm}$. Upon rescaling, the two pieces take the form
\begin{align}
\notag
\bar\La^\text{CSL}_{\phi}={}&\frac12(g^{\mu\nu}_\parallel+v^2g^{\mu\nu}_\perp)\bar\de_\mu\phi\bar\de_\nu\phi-\frac12\phi^2\cos\phi_0,\\
\bar\La^\text{CSL}_{n_\pm}={}&(g^{\mu\nu}_\parallel+v^2g^{\mu\nu}_\perp)\bar\de_\mu n_+\bar\de_\nu n_-+\imag x(n_+\bar\de_0n_--n_-\bar\de_0n_+)\\
\notag
&+n_+n_-\bigl[x^2+(\phi_0')^2-\cos\phi_0-2x\bar H\phi_0'].
\end{align}

The bilinear Lagrangian $\La^\text{CSL}_{\phi}$ of the neutral pion sector is, apart from the appearance of the velocity parameter $v$, identical to that previously found in QCD~\cite{Brauner:2016pko}, and we can therefore merely quote the result for the spectrum. It consists of two energy bands, the lower of which is gapless in accord with the fact that the CSL state spontaneously breaks translations in the direction of the magnetic field. In the gapless (valence) band, the transverse motion is pseudo-relativistic with phase velocity $v$. The dispersion relation is, however, nonlinear in the longitudinal direction, that is, along the magnetic field. The phase velocity at long wavelengths is given by eq.~(5.3) of ref.~\cite{Brauner:2016pko}. Obviously, no instability is present in the neutral pion sector.

The $\La^\text{CSL}_{n_\pm}$ part of the bilinear Lagrangian describes the propagation of (anti)diquarks. Upon Fourier transform in time and the transverse coordinates, one readily finds that their (dimensionless) dispersion relations can be expressed as
\begin{equation}
\bar\omega(\bar{\vek p})=\sqrt{v^2\bar{\vek p}_\perp^2+\lambda}\mp x,
\label{diquarkdisp}
\end{equation}
where $\lambda$ runs over the eigenvalues of the differential operator
\begin{equation}
\Delta\equiv-\bar\de_z^2-(\phi_0')^2+\cos\phi_0+2x\bar H\phi_0'.
\end{equation}
Upon a further rescaling of the coordinate, $\bar{\bar z}\equiv\bar z/k$, using eqs.~\eqref{CSLsolution} and \eqref{CSLvacuum} and the properties of the Jacobi elliptic functions, the eigenvalue problem for $\Delta$ can be recast as
\begin{equation}
\lambda=\frac1{k^2}(\tilde\lambda-4-k^2),
\label{lambda}
\end{equation}
where $\tilde\lambda$ runs over the eigenvalues of another operator,
\begin{equation}
\tilde\Delta\equiv-\bar{\bar\de}_z^2+6k^2\sn^2(\bar{\bar z},k)+\frac{16}\pi E(k)\dn(\bar{\bar z},k)\equiv-\bar{\bar\de}_z^2+V(\bar{\bar z},k),
\label{Deltatilde}
\end{equation}
$\dn$ being another of the Jacobi elliptic functions.

In order to see whether there is a region in the CSL phase where the diquark mode becomes unstable, we have to check whether for some values of $x$ and $\bar H$, the smallest eigenvalue $\lambda$ drops down to $x^2$. The operator $\tilde\Delta$, and thus also the eigenvalues $\lambda$, depends only on the elliptic modulus $k$. The most convenient way to find the condition for the (in)stability of the diquark on the CSL background is therefore to fix the value of $x$, find the $k$ for which the smallest eigenvalue of $\Delta$ equals $x^2$, and then determine the magnetic field from eq.~\eqref{CSLvacuum}.

\begin{figure}
\begin{center}
\includegraphics[scale=1]{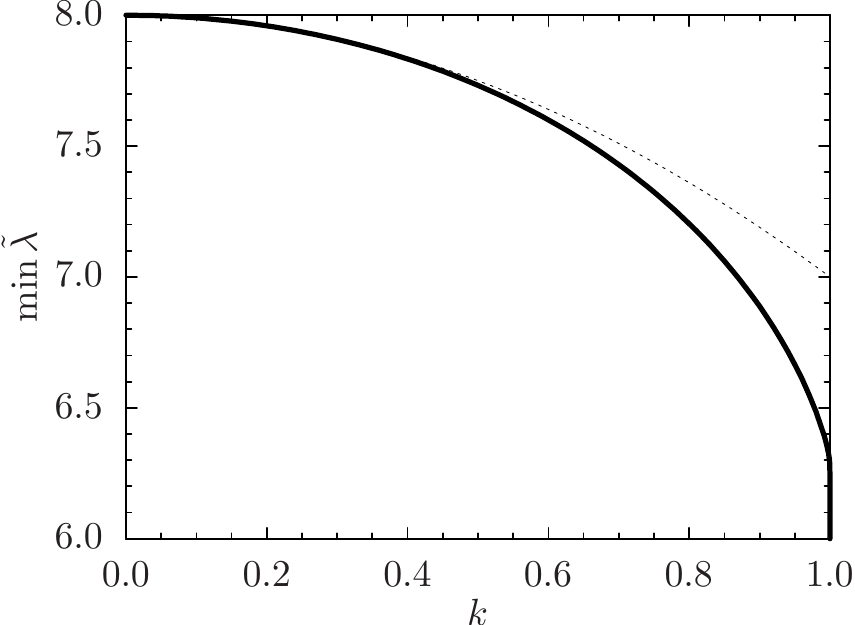}
\end{center}
\caption{The smallest eigenvalue (ground state) of the operator $\tilde\Delta$, defined by eq.~\eqref{Deltatilde}, as a function of $k$. The dashed line indicates the asymptotic behavior for small $k$ given by eq.~\eqref{lambdasmallk}.}
\label{fig:lambda}
\end{figure}

It remains to find the ground state, that is the lowest eigenvalue, of $\tilde\Delta$ as a function of $k$. This can be done numerically using the variational principle, and the results are shown in figure~\ref{fig:lambda}. Some analytic insight into the results can be achieved by looking at the asymptotic behavior of the potential $V(\bar{\bar z},k)$ as a function of $k$. First, we have $\lim\limits_{k\to0+}V(\bar{\bar z},k)=8$. Thus, at $k=0$, the operator $\tilde\Delta$ describes a free particle with spectrum starting at $\min\tilde\lambda=8$. Second, we find that
\begin{equation}
V(\bar{\bar z},1)=6\tanh^2\bar{\bar z}+\frac{16}\pi\frac1{\cosh\bar{\bar z}}.
\label{Vkeq1}
\end{equation}
Remarkably, this potential does not have any bound state in spite of its minimum at the origin.\footnote{This was verified by a numerical examination of the spectrum of the one-parameter set of potentials $V_\alpha(\bar{\bar z})\equiv6\tanh^2\bar{\bar z}+16\alpha/(\pi\cosh\bar{\bar z})$. For $\alpha=0$, this is equivalent to the well-known P\"oschl-Teller potential, whose spectrum can be found analytically. It consists of two bound states with eigenvalues 2 and 5, and a continuous spectrum above 6. For $0<\alpha<1$, the spectrum was investigated numerically using the variational method. With increasing $\alpha$, the eigenvalues corresponding to the bound states grow and eventually disappear in the continuum: the upper bound state disappears for $\alpha\approx0.4$, the lower bound state later at $\alpha\approx0.9$. Hence for $\alpha=1$, corresponding to the potential~\eqref{Vkeq1}, there is no bound state left.} Accordingly, it has a single energy band, starting at $\tilde\lambda=6$. We thus conclude that $\lim\limits_{k\to0+}(\min\tilde\lambda)=8$ and $\lim\limits_{k\to1-}(\min\tilde\lambda)=6$.

The analysis for small $k$ can be further refined by considering the first nontrivial order in the series expansion of the potential in powers of $k$,
\begin{equation}
V(\bar{\bar z},k)=8-2k^2\cos^2\bar{\bar z}+\mathcal O(k^4).
\end{equation}
The Schr\"odinger equation with this potential is a special case of the so-called Mathieu equation. Mapping the eigenvalue problem on the known properties of the Mathieu equation, see for instance eqs.~(4.18)--(4.20) in ref.~\cite{Brauner:2017mui}, leads to
\begin{equation}
\tilde\lambda=8-k^2+\mathcal O(k^4).
\label{lambdasmallk}
\end{equation}
This is indicated by the dashed line in figure~\ref{fig:lambda}.

Using the numerical results displayed in figure~\ref{fig:lambda} leads to the stability limit for diquarks on the CSL background, indicated by the lower dashed line in figure~\ref{fig:phase_diagram}. We observe that the diquark mode only becomes unstable at magnetic fields below the transition to the BEC phase. In other words, the CSL phase as indicated in figure~\ref{fig:phase_diagram} is stable under small fluctuations; the lower dashed line merely denotes the spinodal curve where the CSL state ceases to be a local minimum of the Hamiltonian.

\begin{figure}
\begin{center}
\includegraphics[scale=1]{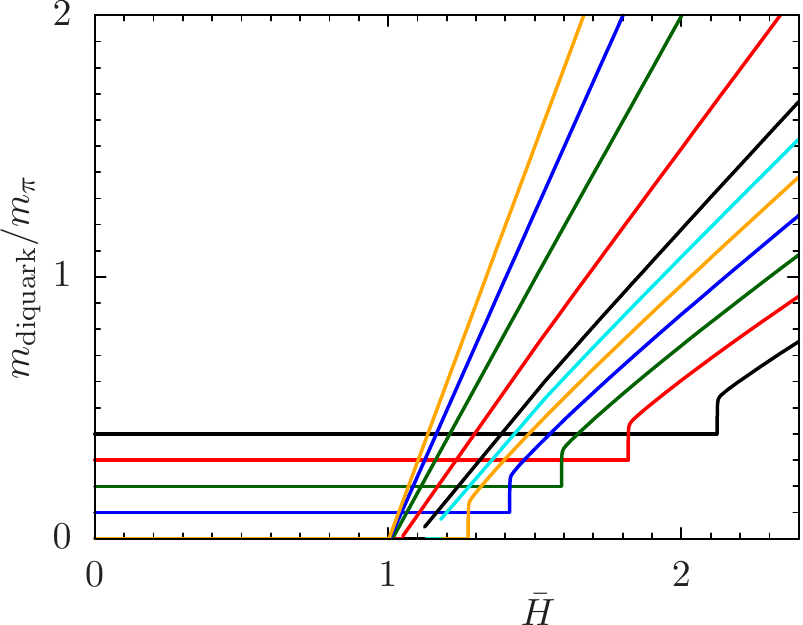}
\caption{Rescaled mass of the lighter of the two diquarks as a function of the magnetic field $\bar H$ for several fixed values of the rescaled chemical potential $x$. Going from right to the left, the individual curves correspond to $x=0.6$ (black), $0.7$ (red), $0.8$ (green), $0.9$ (blue), $1.0$ (orange), $1.1$ (cyan), $1.2$ (black), $1.5$ (red), $2.0$ (green), $2.5$ (blue), and $3.0$ (orange). The first-order nature of the transition between the BEC and CSL phases is clearly visible for the $x=1.1$ and $x=1.2$ curves. For higher $x$, the discontinuity is so small that it is not visible in the graph.}
\label{fig:diquark}
\end{center}
\end{figure}

The dispersion relation~\eqref{diquarkdisp} combined with the data shown in figure~\ref{fig:lambda} can be used without delving further into the spectrum of the operator $\Delta$ to determine the mass spectrum of diquarks. Indeed, by eq.~\eqref{diquarkdisp}, the (rescaled) mass, that is energy at zero momentum, of the diquark-antiquark pair equals $\min\lambda\mp x$. In figure~\ref{fig:diquark}, we show the mass of the lighter of the two states as a function of magnetic field for several fixed values of the baryon chemical potential. This further highlights the location of the transition between the BEC and CSL phases.

\section{Summary and conclusions}
\label{sec:summary}

In this paper, we have analyzed in detail the effect of chiral anomaly on the phase diagram of (pseudo)real QCD-like theories subject to a strong external magnetic field. We showed that under fairly general assumptions, their phase diagram features a nonuniform phase of the CSL type, where nonzero baryon density is generated by a topological crystalline condensate of neutral pions, similarly to what was previously observed in QCD~\cite{Brauner:2016pko}. All the important results were discussed in detail in the main text of the paper, here we therefore merely append a few concluding remarks.

First, as already explained in the prequel to this paper~\cite{Brauner:2019rjg}, the main result as displayed in figure~\ref{fig:phase_diagram} should be taken with a grain of salt. Namely, all the parameters of the effective Lagrangian~\eqref{lagrangian} are given by a priori unknown functions of the magnetic field. Given that, for instance, $\bar H=CH/[f_\pi(H)]^2$ by eq.~\eqref{barH}, the variables used to label the axes in figure~\ref{fig:phase_diagram} represent a complicated nonlinear mapping of the physical parameters $\mu_B$ and $H$. It is therefore not a priori obvious whether the region $\bar H>1$, where the inhomogeneous CSL phase appears, can in fact be reached for any finite value of $H$.

Based on eq.~\eqref{Hcrit}, one can argue that in theories with sufficiently large gauge group and its sufficiently large representation (that is large $d$), the critical magnetic field for the transition to the CSL phase is small and thus under theoretical control. Indeed, the condition $\bar H>1$ can be rewritten as $H/(4\pi f_\pi)^2>1/(2d)$. Combined with the fact that $4\pi f_\pi$ is the loop factor that controls the derivative expansion of the EFT~\cite{Manohar:1983md}, this suggests that the critical field for the formation of the CSL is within the reach of the EFT. Using an explicit one-loop expression for the $H$-dependence of $f_\pi$, available in the literature, shows that $f_\pi$ can, in fact, be treated as a constant; the correction to the critical magnetic field ensuing from its $H$-dependence is suppressed by $1/d$ and thus negligible for large $d$~\cite{Brauner:2019rjg}.

For theories with small $d$, most notably for two-color QCD, our results are somewhat inconclusive though. In order to decide whether the $\bar H>1$ region can be reached with a finite magnetic field, one would need detailed input on the magnetic field dependence of the pion decay constant $f_\pi(H)$. Such an input is, to the best of our knowledge, currently not available.

Second, by focusing solely on theories free of the sign problem, we made a prediction for the phase diagram of a class of QCD-like theories that can serve as a reference for future lattice simulations of inhomogeneous phases of dense quark matter. On the lattice, the presence of the CSL phase can in principle be tested in several different, more or less explicit ways. The most straightforward, but also most brute force, approach would be to look for the nonuniform order, either by trying to study directly the order parameter, or its correlator at long distances~\cite{Yamamoto:2014lia}. Alternatively, one could supplement the EFT analysis presented here by a nonperturbative evaluation of the neutral pion decay constant as explained above. Last but not least, the presence of the characteristic roton-like mode in the spectrum, as displayed in figures~\ref{fig:NGdisp} and~\ref{fig:dispersions}, makes it possible to demonstrate the instability towards the formation of an inhomogeneous ground state by studying detailed properties of the adjacent homogeneous BEC phase.\footnote{The utility of the roton-like dispersion relation as a smoking gun that hints at a presence of a nonuniform phase was pointed out to us by Gergely Endr\H odi.}

Finally, although we did focus solely on theories free of the sign problem, CSL is clearly a much more general phenomenon, only requiring an anomalous coupling of an electrically neutral degree of freedom to an external magnetic field and a chemical potential. Apart from QCD at nonzero baryon chemical potential studied in ref.~\cite{Brauner:2016pko}, this applies in particular to QCD at nonzero \emph{isospin} chemical potential $\mu_I$. This theory is known to be free of the sign problem in a vanishing magnetic field~\cite{Son:2000xc},\footnote{Switching on a magnetic field destroys the positivity of the determinant of the Dirac operator unless one tunes the quark electric charges so that $q_u=q_d$~\cite{Endrodi:2014lja}. In this case, all three pions are electrically neutral and the resulting low-energy EFT is equivalent to that studied in this paper up to a replacement of the (anti)diquark degrees of freedom with the ``charged'' pions $\pi^\pm$ and of $\mu_B$ with $\mu_I$. The only difference is in the value of the coefficient $C$; here eq.~\eqref{Cdef} should be replaced with $C=\frac d{8\pi^2}(q_u+q_d)$.} and its phase diagram has been studied thoroughly with an increasing precision, using both lattice simulations~\cite{Brandt:2017oyy} and EFT~\cite{Adhikari:2019mdk}. By adding a gauge field for isospin to the derivation of the WZ term in section~\ref{subsec:WZ}, one can see that an external magnetic field in combination with the isospin chemical potential can also give rise to the CSL phase. The resulting low-energy EFT is identical to that analyzed in ref.~\cite{Brauner:2016pko} except for a different overall normalization of the WZ term, fixed by eq.~\eqref{WZterm2}.

A similar comment applies to the class of QCD-like theories covered here but with nonzero quark electric charges that do \emph{not} satisfy the condition $q_u=-q_d$, assumed throughout our paper. In this case, all the diquarks are electrically charged and thus acquire a gap from the magnetic field. The low-energy spectrum then only contains the neutral pion, and the analysis of ref.~\cite{Brauner:2016pko} applies without modification, possibly only with a different coefficient of the WZ term.


\acknowledgments

The present work was inspired by a discussion with Hiromichi Nishimura. We are indebted to Thomas Cohen, Gergely Endr\H{o}di, Philippe de Forcrand, Simon Hands, Carlos Hoyos, Aleksi Kurkela, Eugenio Meg\'{\i}as, Andreas Schmitt, Igor Shovkovy, and especially to Naoki Yamamoto, for insightful discussions and comments. This work has been supported by a ToppForsk-UiS grant no.~PR-10614.


\appendix

\section{Variational minimization of the Hamiltonian}
\label{app:minimization}

The ground state of the EFT is given by the absolute minimum of (the spatial average of) the Hamiltonian~\eqref{HamDimless}. Unfortunately, unlike in QCD itself~\cite{Brauner:2016pko}, it is not straightforward to carry out the minimization due to the presence of the diquark degrees of freedom, and it is not even obvious whether a closed analytic expression for the ground state exists at all. In this appendix, we report on some of the checks we have done to convince ourselves that no state of lower energy than those analyzed in section~\ref{sec:away} exists, and thus figure~\ref{fig:phase_diagram} is the correct phase diagram of the EFT.

To that end, we used the variational principle with several different analytic ans\"atze, which may contain the known BEC and CSL states as special cases. These are described in some detail in the following subsections.


\subsection{States with constant $\theta$}

In section~\ref{sec:away}, we saw that it is straightforward to minimize the Hamiltonian on the subspaces of uniform states (leading to the BEC phase), and of states with $\theta=0$ (leading to the CSL phase). In fact, both of these can be embedded in a larger class of trial states, for which a fully analytic solution still exists, namely those where $\theta$ is assumed to be constant but possibly nonzero. The Hamiltonian~\eqref{HamDimless} then reduces to $\bar\Ha=\bar\Ha_\phi+\bar\Ha_\theta$, where
\begin{equation}
\bar\Ha_\phi\equiv\frac12(\phi')^2\cos^2\theta-\cos\theta\cos\phi-x\bar H\phi'\cos^2\theta,\qquad
\bar\Ha_\theta\equiv-\frac{x^2}2\sin^2\theta+1,
\end{equation}
and we already set $\alpha=0$ without loss of generality. The total energy resulting from $\bar\Ha_\phi$ can be minimized in the same way as in ref.~\cite{Brauner:2016pko}. Setting $c\equiv\cos\theta$, one finds that the stationary states of the Hamiltonian are given by a straightforward generalization of eq.~\eqref{CSLsolution},
\begin{equation}
\cos\frac{\langle\phi(\bar z)\rangle}2=\sn\left(\frac{\bar z}{k\sqrt c},k\right).
\label{SolCSLlike}
\end{equation}
The spatially averaged energy density of the solution equals
\begin{equation}
\bar\Ha_\phi=\frac{4cE(k)}{k^2K(k)}+c\left(1-\frac2{k^2}\right)-\frac{\pi x\bar H c^{3/2}}{kK(k)},
\end{equation}
where $K(k)$ is the complete elliptic integral of the first kind. All that is left to do is to minimize this with respect to the elliptic modulus $k$ and add the resulting energy to $\bar\Ha_\theta$. We thus arrive at an effective Hamiltonian as a function of $\theta$ alone,
\begin{equation}
\bar\Ha(\theta)=-\frac{x^2}2\sin^2\theta+1+\biggl[1-\frac2{k(\theta)^2}\biggr]\cos\theta,
\label{Hefftheta}
\end{equation}
where the function $k(\theta)$ is defined implicitly by a condition, generalizing eq.~\eqref{CSLvacuum},
\begin{equation}
\frac{E(k)}k=\frac{\pi x\bar H\sqrt{\cos\theta}}4.
\end{equation}
This equation has a unique solution for $k$ only for $\cos\theta>(4/\pi x\bar H)^2$. For values of $\theta$ not satisfying this condition, $k=1$ is to be used in the Hamiltonian~\eqref{Hefftheta}.

It remains to minimize $\bar\Ha(\theta)$ as defined by eq.~\eqref{Hefftheta} with respect to $\theta$ for given values of $x$ and $\bar H$. As it turns out, however, the ground state has, for any $x$ and $\bar H$, either $\phi=0$ (BEC state) or $\theta=0$ (CSL state). Within the class of states considered here, there is no other state of even lower energy.


\subsection{States with periodic $\theta$ and linear $\phi$}
\label{ApLinPh}

Our first ansatz discussed above is probably not the most natural one. After all, once translational invariance is spontaneously broken by a spatially varying $\phi$ as in the CSL state, one would expect $\theta$, if nonzero, to be nonuniform as well. In this case, we have to make some further assumptions on the spatial profile of both $\theta$ and $\phi$ in order to be able to compute, and minimize, the energy quasi-analytically.

To that end, we first recall the physical meaning of the angular variables $\theta$ and $\phi$ as introduced by eq.~\eqref{angle}. Since we can set $\alpha=0$ without loss of generality, $\theta$ and $\phi$ are nothing but the usual spherical coordinates parameterizing a unit two-sphere in the $n_0$-$n_1$-$n_3$ space. The angle $\theta$ measures the azimuthal deviation from the equator, lying in the $n_0$-$n_3$ plane, whereas $\phi$ is the usual polar angle. The BEC state maps to a fixed point in the $n_0$-$n_1$ plane, whereas the CSL state describes a periodic motion along the equator with the $z$-coordinate playing the role of ``time.''

A nonuniform state that should have a lower energy than the uniform BEC state, must make use of the anomalous contribution to the Hamiltonian~\eqref{HamDimless}, proportional to $\phi'$, and thus have nontrivial topology due to winding in the $\phi$-direction. We want to see if it is possible to lower the energy as compared to the CSL state by allowing the azimuthal angle $\theta$ to vary as well. Due to the nonperiodic nature of the azimuthal angle, we expect $\theta$ to oscillate within some finite range.

The simplest ansatz with these properties takes the form
\begin{equation}
\theta(\bar z)=\theta_0+K\sin\left(\frac{2\pi}{l}\bar z\right),\qquad
\phi(\bar z)=\phi_0+\frac{2\pi}{l}\bar z,
\label{ans2}
\end{equation}
where $\theta_0$, $\phi_0$, $K$ and $l$ are parameters. The linear ansatz for $\phi$ is motivated by the profile of the CSL state in the chiral limit~\eqref{CSLchiral}, which is combined with the simplest ansatz for $\theta$ oscillating with the same period as the CSL state. Note that this ansatz contains neither the BEC nor the CSL state as its special cases. First, the spatial profile of $\phi(\bar z)$ is fixed to a simple approximation to the more complicated CSL state. Second, the ansatz~\eqref{ans2} has a nontrivial topology for any finite $l$; the BEC state can be formally recovered by setting $l=\infty$, which however does \emph{not} correspond to the limit of eq.~\eqref{ans2} as $l\to\infty$.

With the ansatz~\eqref{ans2}, the spatial average of the dimensionless Hamiltonian~\eqref{HamDimless} can be worked out analytically,
\begin{equation}
\begin{split}
\bar\Ha={}&\frac{K^2}{4}\left(\frac{2\pi}{l}\right)^2+\frac{1}{4}\biggl[\left(\frac{2\pi}{l}\right)^2-2x\bar{H}\frac{2\pi}{l}-x^2\biggr]+1\\
&+\frac{1}{4}\biggl[\left(\frac{2\pi}{l}\right)^2-2x\bar{H}\frac{2\pi}{l}+x^2\biggr]J_0(2K)\cos(2\theta_0)-J_1(K)\sin\theta_0\sin\phi_0,
\end{split}
\label{HeffAns2}
\end{equation}
where $J_n$ are the Bessel functions of the first kind. The function~\eqref{HeffAns2} of the variables $\theta_0$, $\phi_0$, $K$ and $l$ can be systematically minimized using standard methods of multivariate calculus; for some of the local extrema the minimization over one of the variables has to be performed numerically. It turns out, however, that the lowest energy achievable with the ansatz~\eqref{ans2} is not lower than the energy of the ground state found in section~\ref{sec:away} for any values of $x$ and $\bar H$.

Let us remark for completeness that for sufficiently strong magnetic fields, the global minimum of the averaged energy density~\eqref{HeffAns2} is reached by a CSL-like state with $\theta_0=K=0$. For lower magnetic fields, the minimum appears at $\theta_0=\phi_0=\pi/2$ and 
\begin{equation}
\frac{2\pi}{l}=\frac{x\bar{H}\left[1-J_0(2K)\right]}{K^2+1-J_0(2K)};
\end{equation}
the optimum value of $K$ has to be found numerically.


\subsection{States with periodic $\theta$ and CSL-like $\phi$}

Using the ansatz~\eqref{ans2}, it is possible to get quite close to the energy of the CSL state~\eqref{CSLsolution}, but the energy nevertheless stays above the energy thereof. One might blame this on the far-too-simple spatial profile of $\phi(\bar z)$ in eq.~\eqref{ans2}, which would suggest that a more flexible variational ansatz might allow us to further lower the energy.

To that end, recall that the linear ansatz for $\phi(\bar z)$ in eq.~\eqref{ans2} was based on the solution in the chiral limit, which corresponds to small $k$. The full CSL solution~\eqref{CSLsolution} can be accordingly expanded in small $k$; including the first correction beyond the linear contribution gives~\cite{Abramowitz:1972}
\begin{equation}
\phi(\bar z)=-\pi+\frac{2}{k}\left[\left(1-\frac{k^2}{4}\right)\bar z+\frac{k^2}{8}\sin\frac{2\bar z}{k}\right]+\mathcal O(k^2).
\end{equation}
Motivated by the functional dependence of $\phi$ on $\bar z$ in this approximation, we improve on eq.~\eqref{ans2} by making the following ansatz,\footnote{We checked numerically that using this ansatz in QCD, one can approach the exact ground state energy (from above) with $\mathcal{O}(10^{-2})$ precision even for points in the $x$-$\bar{H}$ plane corresponding to $k$ as big as $0.9$. For $k\lesssim0.5$, the $\mathcal{O}(10^{-4})$ precision is reached.}
\begin{equation}
\theta(\bar z)=\theta_0+K_\theta\sin\left(\frac{2\pi}{l_\theta}\bar z+\theta_1\right),\qquad
\phi(\bar z)=\phi_0+\frac{2\pi}{l_\phi}\bar z+K_\phi\sin\left(\frac{2\pi}{l_\phi}\bar z\right),
\label{ans1}
\end{equation}
where $\theta_0$, $\theta_1$, $\phi_0$, $K_\theta$, $K_\phi$, $l_\theta$ and $l_\phi$ are variational parameters. It is assumed that the periods of the spatial variation of $\theta(\bar z)$ and $\phi(\bar z)$ are commensurate, that is, $l_\theta/l_\phi$ is a rational number.

The spatial average of the dimensionless Hamiltonian~\eqref{HamDimless} can still be worked out in a closed form, and reads
\begin{align}
\notag
\bar\Ha={}&\frac{K_\theta^2}{4}\left(\frac{2\pi}{l_\theta}\right)^2+\frac{1}{4}\biggl[\left(\frac{2\pi}{l_\phi}\right)^2-2x\bar{H}\frac{2\pi}{l_\phi}-x^2\biggr]+\frac{K_\phi^2}{2}\left(\frac{2\pi}{l_\phi}\right)^2I_1(l_\theta/l_\phi,\theta_0,\theta_1,K_\theta,2)\\
\label{HeffAns1}
&+K_\phi\biggl[\left(\frac{2\pi}{l_\phi}\right)^2-x\bar{H}\frac{2\pi}{l_\phi}\biggr]I_1(l_\theta/l_\phi,\theta_0,\theta_1,K_\theta,1)\\
\notag
&+\frac{1}{4}\biggl[\left(\frac{2\pi}{l_\phi}\right)^2-2x\bar{H}\frac{2\pi}{l_\phi}+x^2\biggr]J_0(2|K_\theta|)\cos(2\theta_0)-I_2(l_\theta/l_\phi,\phi_0,\theta_0,\theta_1,K_\phi,K_\theta)+1.
\end{align}
The integrals $I_1$ and $I_2$ are defined by
\begin{align}
\label{defI1}
I_1(l_\theta/l_\phi,\theta_0,\theta_1,K_\theta,\alpha)&\equiv\frac{1}{2\pi}\int_0^{2\pi}\cos^\alpha(m\bar z)\cos^2[\theta_0+K_\theta\sin(n\bar z+\theta_1)]\,\dd\bar z,\\
\notag
I_2(l_\theta/l_\phi,\phi_0,\theta_0,\theta_1,K_\phi,K_\theta)&\equiv\frac{1}{2\pi}\int_0^{2\pi}\cos\left(\phi_0+m\bar z+K_\phi\sin m\bar z\right) \cos^2[\theta_0+K_\theta\sin(n\bar z+\theta_1)]\,\dd\bar z,
\end{align}
where $m$, $n$ are coprime integers such as $m/n=l_\theta/l_\phi$.

The spatially averaged energy functional was subsequently minimized numerically with respect to the parameters $\theta_0$, $\theta_1$, $\phi_0$, $K_\theta$, $K_\phi$ and $l_\phi$ for an array of points in the $x$-$\bar{H}$ plane and $l_\theta/l_\phi\in \{1/4,1/3,1/2,2/3,1,3/2,2,3,4\}$. For large enough $\bar{H}$, the minimum was found to be CSL-like, that is with $\theta_0\to 0$ and $K_\theta\to 0$. For lower values of $\bar{H}$, a state with $K_\phi\to 0$ and $l_\theta=l_\phi$ (which is neither CSL-like nor BEC-like) was preferred. The latter corresponds to the trial state investigated in section~\ref{ApLinPh}. No field configuration with energy lower than that of the states analyzed in section~\ref{sec:away} was found for any combination of $x$ and $\bar H$.


\bibliographystyle{JHEP}
\bibliography{references}

\end{document}